\documentclass[aps,prb,twocolumn,amsmath,amssymb,nofootinbib,superscriptaddress,floatfix,reprint,longbibliography]{revtex4-2}
\usepackage[dvips]{graphicx}
\usepackage{latexsym}
\usepackage{amsmath}
\usepackage{amsfonts}
\usepackage{amssymb}
\usepackage{bm}
\usepackage{color}
\usepackage{colortbl}
\usepackage{txfonts}
\usepackage{float}
\usepackage{url}
\usepackage[colorlinks=true, urlcolor=blue, linkcolor=blue, citecolor=blue, pdftex]{hyperref}
\usepackage{ulem}
\usepackage{multirow}
\usepackage{makecell}

\begin{document}

\title {Josephson diodes induced by the loop current states}

\author{Qi-Kai Shen}

\affiliation{Department of Physics and Institute for Quantum Science and Technology, Shanghai University, Shanghai 200444, China}
\author{Yi Zhang}
\email{zhangyi821@shu.edu.cn}
\affiliation{Department of Physics and Institute for Quantum Science and Technology, Shanghai University, Shanghai 200444, China}
\affiliation{Shanghai Key Laboratory of High Temperature Superconductors and International Center of Quantum and Molecular Structures, Shanghai University, Shanghai 200444, China}  

\date{\today}

\begin{abstract}
We study the diode effect of the supercurrent in the Josephson junctions with the loop current states as the tunneling barrier.
The loop current states are realized by the Haldane model which preserves the inversion symmetry and thus forbids the diode effect. We demonstrate how the inversion symmetry can be broken in the monolayer and bilayer systems. In the monolayer system, inversion symmetry can be broken by either introducing a sublattice staggered potential for the Haldane model or introducing a modified Haldane model, and in the bilayer system, it can be broken by either staking the two layers with opposite current directions or by directly applying an electric field perpendicular to the layers. We further show that the diode efficiency can be tuned by the interlayer coupling and the strength of the electric field or interlayer voltage.
Our results provide another route to realize the Josephson diode effect by breaking the time-reversal symmetry through the loop-current states.
\end{abstract}


\maketitle

\section{Introduction}
Superconducting diode effect (SDE) refers to the phenomenon where critical supercurrents flowing in opposite directions have different magnitudes. This non-reciprocal transport effect in the superconducting system can play a similar role as the diodes formed by the p-n junctions in the semiconductor industry and has potential applications in the development of dissipationless electronics~\cite{sde1,sde2}. 

SDE has been observed experimentally in numerous superconducting systems~\cite{ando,sde_NbSe2,kim_2023,tTLG,cvs_2024,ryohei_2017,Miyasaka_2021,enze_2020,lyu_2021,narita_2022,Stra_2022}. Such diode effect is also realized in the various Josephson junction (JJ) systems~\cite{ali,Bau_2022,Baumgartner_2022,NiTe2,tbg,gupta_2023,bianca_2022,chen2024,ghosh2024hightemperature,lotfizadeh2024superconducting,kim2024intrinsic}, which is named the Josephson diode effect (JDE). It provides an extra knob to realize and tune the diode effect by designing the proper barrier layer of the JJs.   

On the side of theoretical development, JDE was first proposed based on the electron and hole-doped superconductors (SC) in analog to the semiconducting p-n junctions~\cite{hu}. 
From the symmetry point of view, both SDE and JDE require the system to break both inversion and time-reversal symmetry.
These special JJs with both symmetry broken were intensively studied~\cite{buzdin_2008,huxiao_2014,yuli_2014,julia_2015,kou_2016,Marco_2019,phi0_2013,phi0_2016,phi0_2017,phi0_2018,phi0_2020,phi0_2021,phi0_2022}. One class of such JJs is called Josephson $\phi_0$ junction, which provides a possible mechanism to realize the JDE.  
More recently, as stimulated by the experimental observations, numerous theoretical proposals~\cite{nagaosa1,hejun,yuan,yanse,fu_2022,tTLGt,yi2022,ruben_2022,Kara_2022,law_2023,lu_2023,alv_2023,naka_2024,wang2024efficient,cayao2024enhancing,soori2024josephson,volkov2024josephson,cuozzo2024microwavetunable,seoanesouto2024tuning,debnath2024gatetunable,meyer2024josephson,fracassi2024anomalous,zazunov2024nonreciprocal,huang2024superconducting,karabassov2024anisotropic,correa2024theory,greco2024double,marg_2023,mar_2023,fukaya_2024,fu_2024,fu_arxiv} have been suggested for these effects.
Moreover, the diode effect for the spin transport in the JJs is also proposed~\cite{yi2023,sun_2024}. 

While most of the realizations of the SDE/JDE require a finite external magnetic field or internal net magnetic moment to break the time-reversal symmetry, recent observations of the field-free diode effect in the JJs~\cite{ali} and SC bulk~\cite{cvs_2024} significantly expand the potential scope for this phenomenon but at the same time call for new theoretical mechanisms to realize it.
Besides the systems with the magnetic field, the loop current states are distinct phases of matter that break the time-reversal symmetry, which are extensively studied in different systems, ranging from the Haldane model~\cite{haldane} on the honeycomb lattice to the various forms of flux phases in the square lattice~\cite{afflex_1988,ubbens_1992,patrick_2006,varma_1997,varma_1999,cha_2001,chetan_2000} and triangular lattice~\cite{ven_2016,liu_2016,classen_2019,yuge_2024}.
The recent discovery of the exotic charge density wave in nonmagnetic AV$_3$Sb$_5$ (A = K, Rb, Cs)~\cite{cdw1,cdw2,cdw3,cdw4,cdw5,cdw6,cdw7,cdw8,vidya_2024} 
provides another possible material realization of the loop current states~\cite{feng_2021,denner_2021,park_2021,lin_2021,gu_2022,dong_2023,wang_2022,Hiroshi_2023,wu_2024}.
In this work, we propose a theoretical model to realize the JDE using the loop current states as the tunneling barrier of the JJ, which serves as the source of the time-reversal symmetry breaking. 
We construct the loop current states in the honeycomb lattice with monolayer and bilayer systems. For the monolayer case, JDE can be realized either by introducing a potential difference for the two sublattices or by modifying the current directions of the Haldane model that breaks the inversion symmetry directly. For the AB-staked bilayer case, JDE can be induced by either stacking the two layers with opposite loop currents or applying an out-of-plane electric field. For the bilayer system, we further analyze the effect of the interlayer coupling and displacement field on the diode efficiency.

This work is organized as follows. In Sec.\ref{sec:tb}, we introduce the tight-binding models that host the loop current order. In Sec.\ref{sec:JJs}, we describe the construction of the Josephson junction within the tight-binding model. The main results, including the current-phase relation of the junction, are presented in Sec.\ref{sec:CPR}, followed by an analysis of the diode efficiency in Sec.\ref{sec:eff}. In Sec.\ref{sec:sym}, we provide a symmetry analysis of the Josephson diode effect (JDE), and in Sec.\ref{sec:dis}, we examine its robustness against disorder. Finally, we conclude our discussion in Sec.~\ref{sec:con}.


\section{Tight-binding Models}
\label{sec:tb}
\begin{figure}
\centering
\includegraphics[width=3.3in]{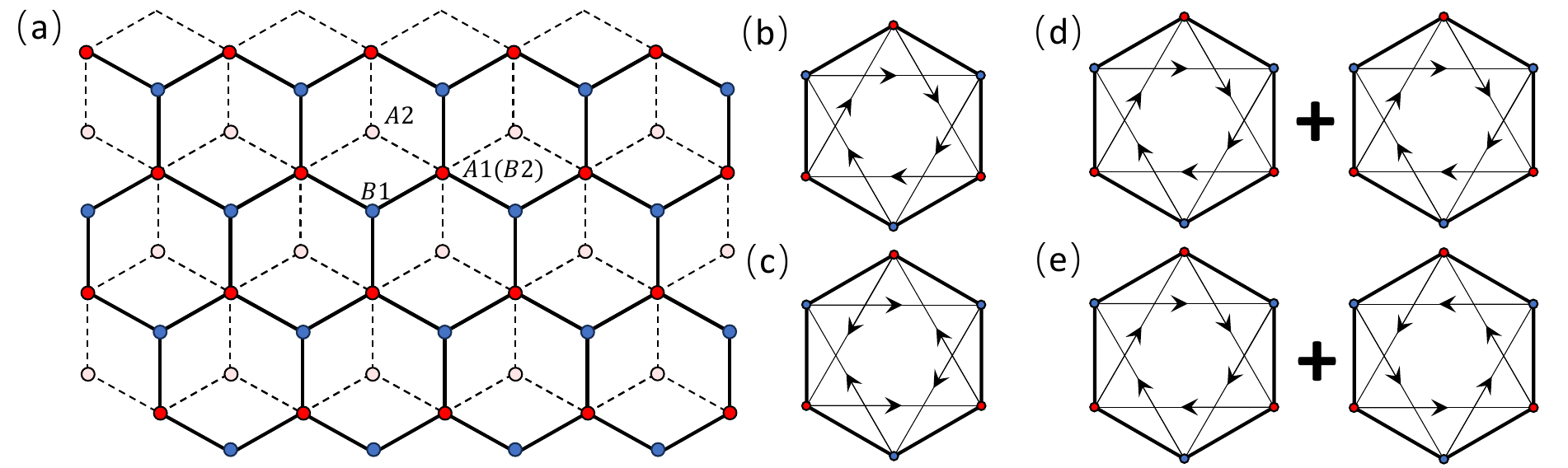}
\caption{(a) Schematic plot of the bilayer honeycomb lattice. The first layer is characterized by atoms labeled as A1 and B1, connected by solid lines, while the second layer is indicated by A2 and B2 atoms, connected by dashed lines. The stacking of A1 atoms from the first layer with B2 atoms from the second layer gives rise to the bilayer HMs. (b) In the HM, the NNN hopping directions of atoms A and B are opposite. (c) In the MHM, the NNN hopping directions of atoms A and B are the same. (d) Two HMs are stacked with the same loop current directions. (e) Two HMs are stacked with opposite loop current directions.}
\label{fig:model}
\end{figure}

As the basic building block of the tunneling barrier of the JJ, we start with the Haldane model (HM)~\cite{haldane} on a honeycomb lattice containing two different types of atoms denoted A and B, as shown by the solid line in Fig.\ref{fig:model}(a). The tight-binding Hamiltonian can be written as
\begin{equation}
    \begin{split}
        H_m=-t\sum_{\left \langle i,j \right\rangle\sigma}c_{i\sigma}^\dagger c_{j\sigma}-
        t_1 \sum_{\left \langle\left \langle i,j \right\rangle\right \rangle\sigma} e^{-i\nu_{ij}\phi} c_{i\sigma}^\dagger c_{j\sigma}
    \end{split}
    \label{con:monolayer_H}
\end{equation}
where $i(j)$ labels the lattice sites and $\sigma$ is the electron spin index. The operators $c_{i\sigma}(c_{i\sigma}^\dagger)$ represent the annihilation (creation) operators of fermions located at site $i$ with spin $\sigma$. Moreover, $t$ is the nearest-neighbor (NN) hopping and $t_1=0.4t$ denotes the amplitude of the next-nearest-neighbor (NNN) hopping. The complex NNN hopping breaks time-reversal symmetry by forming a loop current state, with $\nu_{i,j}=2/\sqrt{3}(\hat{\textbf{d}}_1\times\hat{\textbf{d}}_2)_z=\pm 1$, where $\hat{\textbf{d}}_1$ and $\hat{\textbf{d}}_2$ are the unit vectors along the two bonds that constitute the NNN bonds from site $i$ to site $j$. The signs of the NNN hopping phase are schematically shown in Fig.~\ref{fig:model}(b), which is positive along the arrow direction. From the sign pattern, we can see that the bonds along the same direction, but consisting of different types of atoms, have hoppings with opposite phases, which preserves the inversion symmetry.
Since the inversion symmetry has to be broken to achieve the JDE, we further consider the 
modified Haldane model (MHM)~\cite{colomes2018antichiral}, where the sign of the hopping phase of the bonds consisting of atom B are reversed, as illustrated in Fig.\ref{fig:model}(c). Therefore, in the MHM, both time-reversal and inversion symmetry are broken, which meets the basic symmetry requirement to realize the JDE.

Besides the monolayer system, we also consider the bilayer HMs, where the two layers of the honeycomb lattice are Bernal stacked, with one layer's A atoms stacked with the other layer's B atoms as shown in Fig.\ref{fig:model}(a).
The Hamiltonian for the bilayer system can thus be written as
\begin{equation}
    \begin{split}
        H_b&=H_1+H_2+H_{12}\\
        H_l&=-t\sum_{\left \langle i,j \right\rangle\sigma}c_{il\sigma}^\dagger c_{jl\sigma}-t_1 \sum_{\left \langle\left \langle i,j\right\rangle\right\rangle\sigma} e^{-i\nu_{lij}\phi} c_{il\sigma}^\dagger c_{jl\sigma}\\
        H_{12}&=-t_\perp \sum_{i,\sigma} c_{i,A1 \sigma}^\dagger c_{i,B2 \sigma}+h.c.
    \end{split}
    \label{con:bilayer_H}
\end{equation}
where $l$ is the layer index, $H_l$ represents the Hamiltonian for each layer, and $H_{12}$ corresponds to the coupling between the two layers. Here, we assume that only the atoms stacked on top of each other between the two layers (A1-B2) contribute to this interlayer coupling $t_\perp$. The introduction of the extra layer brings more freedom to manipulate the symmetry of the system. Here, we consider the loop current pattern of each layer forms the original HM, which is demonstrated to be more stable in a mean-field calculation~\cite{zhu2013ordered}.
Then, the remaining degree of freedom is the relative orientation of the loop current of the two layers. When the two layers have the loop current with the same orientation as depicted in Fig.~\ref{fig:model}(d), the inversion symmetry remains preserved and the bilayer system has similar properties as the monolayer HM. In order to break the inversion symmetry in this state, an electric field perpendicular to the layers can be introduced.
On the other hand, when the orientations of the loop current are opposite for each layer as shown in Fig.~\ref{fig:model}(e), the inversion symmetry is broken by having an odd parity with respect to two layers. This is the magnetoelectric state theoretically proposed in the bilayer honeycomb lattice with NNN repulsion~\cite{zhu2013ordered}. 



\begin{figure}
\centering
\includegraphics[width=3in]{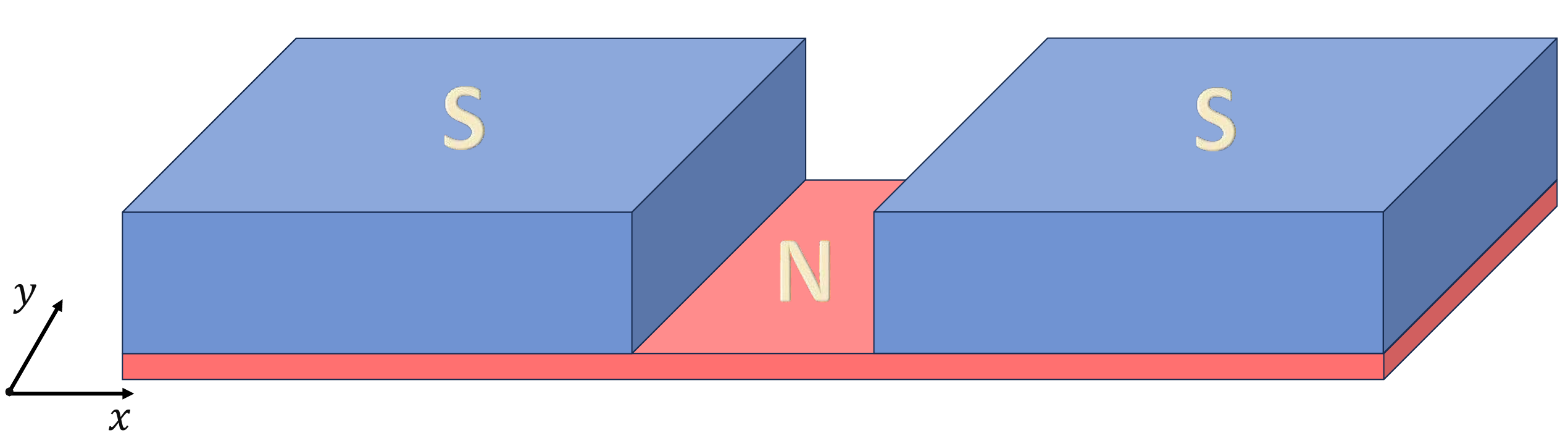}
\caption{Schematic of a planar Josephson junction in the $x-y$ plane. The blue region signifies the superconducting region (S), while the red region represents the normal region (N). The periodic boundary condition is used for the y-direction and the Josephson current flows in the x-direction.}
\label{fig:Joseogson_Diode}
\end{figure}

\section{Construction of the Josephson junction}
\label{sec:JJs}
We next construct the planar Josephson junction using the monolayer and bilayer systems as the tunneling barrier.
As depicted in Fig.\ref{fig:Joseogson_Diode}, this can be achieved by placing two s-wave superconductors (blue) on a two-dimensional material with loop current order (red). Due to the proximity effect, the two-dimensional material underneath the superconductors becomes superconducting. The region without superconductors in between remains in the normal state, thus forming a Josephson junction. Since the translation symmetry is preserved in the y-direction, we adopt the periodic boundary condition in this direction. In the y-direction, we consider unit cells $N_y=100$, while in the x-direction, we have 20 layers for the left superconducting region, 2 layers for the middle normal region, and 20 layers for the right superconducting region.

The Hamiltonian of this Josephson junction, structured upon a honeycomb lattice, can be expressed as follows:
\begin{equation}
    \begin{split}
        H&=H_{SL}+H_{SN}+H_{N}+H_{NS}+H_{SR}\\
        H_{SL}&=H_{0}+\Delta e^{i\varphi} \sum_i c_{i\uparrow}^\dagger c_{i\downarrow}^\dagger+h.c.-\mu_0 \sum_{i,\sigma} c_{i \sigma}^\dagger c_{i \sigma}\\
        H_N&=H_{0}-\mu_0 \sum_{i,\sigma} c_{i \sigma}^\dagger c_{i \sigma}\\
        H_{SR}&=H_{0}+\Delta \sum_i c_{i\uparrow}^\dagger c_{i\downarrow}^\dagger+h.c.-\mu_0 \sum_{i,\sigma} c_{i \sigma}^\dagger c_{i \sigma}
    \end{split}
    \label{eq:jd_hamilton}
\end{equation}
where $H_{0}$ represents the Hamiltonian of monolayer or bilayer system, as derived from either Eq.\ref{con:monolayer_H} or Eq.\ref{con:bilayer_H}. $H_{SL}$ and $H_{SR}$ are the Hamiltonians of the superconducting regions. $H_N$ represents the Hamiltonian of the non-superconducting region. $H_{SN}$ and $H_{NS}$ represent the spin-conserving hoppings between the superconducting and normal sections which include both the NN and NNN hoppings and take the same values and patterns as those in the normal region. The superconducting phase bias for the JJ is assigned to the left superconducting region as $\varphi$ so that the phase in the right superconducting region is set to $0$. The amplitude of the order parameter is taken as $\Delta=0.2t$ for all calculations in this paper.   $\mu_0$ represents the chemical potential of the entire material, which is set to -0.4t in the bilayer model, and set to 0 in the monolayer system. More detailed setups for the tight-binding model of the JJs are provided in the Supplemental Material~\cite{sm}.

The Josephson current flows in the x-direction and the supercurrent through the junction is related to the total energy of the system by\cite{beenakker1991universal}
\begin{equation}
    \begin{split}
        I(\varphi)=\frac{2e}{\hbar}\partial_\varphi \sum_n f(\epsilon_n(\varphi))\epsilon_n(\varphi)
    \end{split}
    \label{con:Equ_I}
\end{equation}
where $\epsilon_n$ represents the $n$th eigenvalue of the total Hamiltonian $H$ and $f(x)$ denotes the Fermi-Dirac distribution function, with the temperature set to be $10^{-3}t$ through the whole paper.
In Eq.~\ref{con:Equ_I}, while considering the contribution of the superconducting current, all states are included.


\section{The current-phase relationship}
\label{sec:CPR}
\begin{figure}
\centering
\includegraphics[width=3.5in]{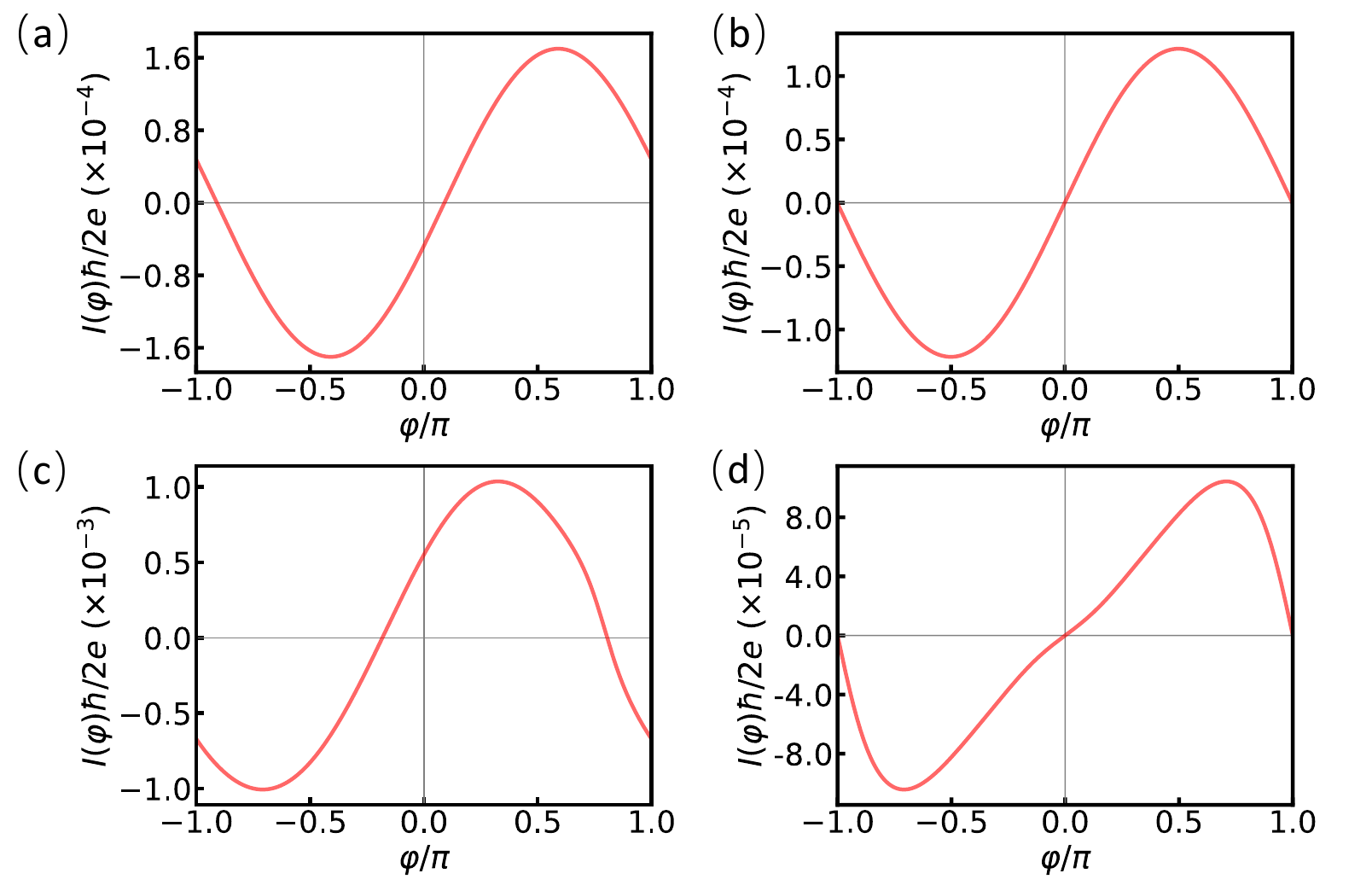}
\caption{The CPR in the monolayer honeycomb lattice. The JJ is constructed by the HM with staggered lattice potential $V_s$ for (a) and (b), and constructed by the MHM for (c) and (d).
(a) and (c) represent the zigzag JJ, which exhibits the diode effect. In contrast, (b) and (d) represent the armchair JJ, which does not show any diode effect. Parameters are taken as $t=1$, $t_1=0.4t$, $\phi=\pi/4$, $V_s=0.2t$, $\Delta=0.2t$, $\mu_0=0$}.
\label{fig:monolayer_I}
\end{figure}

We construct the JJs with both the monolayer and bilayer systems as the tunneling barrier.
Depending on the orientation of the junction, we name the JJ as a zigzag (armchair) JJ, if the edge of the junction along the current flow direction has a zigzag (armchair) shape.
For the monolayer case, when the tunneling barrier is constructed with the HM, the current-phase relation (CPR) shows no diode effect, where the critical current for the positive and negative directions have the same magnitude~\cite{sm}. This is expected since the loop current pattern in the HM preserves the inversion symmetry, which precludes the emergence of the JDE.
One way to break this inversion symmetry is to introduce a staggered lattice potential within the unit cell of HM, which can be described by an extra term
\begin{equation}
    H_s=V_s\sum_{i,\sigma} (c_{i,A \sigma}^\dagger c_{i,A \sigma}-c_{i,B \sigma}^\dagger c_{i,B \sigma}) \ .
\end{equation}
This term is added to Eq.~\ref{con:monolayer_H} which enters the Hamiltonian of the JJ as $H_0$ in Eq.~\ref{eq:jd_hamilton}. The staggered potential can be realized by placing the material on a substrate with a honeycomb structure, such as Hexagonal boron nitride (h-BN), where the two sublattices consist of different atoms.
The CPR for both types of the JJ is shown in Fig.~\ref{fig:monolayer_I}(a, b). Indeed, we can see that the zigzag JJ shows the diode effect where the magnitude of the critical current for both directions differs from each other, i.e., $I_c^+\neq |I_c^-|$ as shown in Fig.~\ref{fig:monolayer_I}(a). Moreover, a finite Josephson current is achieved for the vanishing phase bias $\varphi$, which is directly related to the breaking of both inversion and time-reversal symmetry in this system.
However, the CPR for the armchair JJ does not show any nonreciprocal effect.
We can understand the different behaviors of the two types of JJs from the symmetry argument. From the symmetry point of view, if the system is invariant under a certain operation that also changes the sign of the current operator, the state of the system with a positive current is always connected to another state with its current reversed by this operation. This means the critical current must have the same magnitude for opposite directions, which causes the absence of the JDE.
Since both time-reversal and inversion change the sign of the current operator, the JDE requires both symmetries to be broken. 
However, even in a system with both symmetries broken, as long as there exists a symmetry that changes the sign of the current operator, the JDE is still strictly forbidden. 
The detailed symmetry analysis will be presented in Sec.~\ref{sec:sym}.
On the other hand, if the tunneling barrier is constructed with the MHM, where the inversion symmetry is explicitly broken by the loop current pattern, we also expect the JDE to show up.
As shown in Fig.~\ref{fig:monolayer_I}(c, d), we again find that the JDE appears only for the zigzag JJ due to similar symmetry reasons.


\begin{figure}
\centering
\includegraphics[width=3.5in]{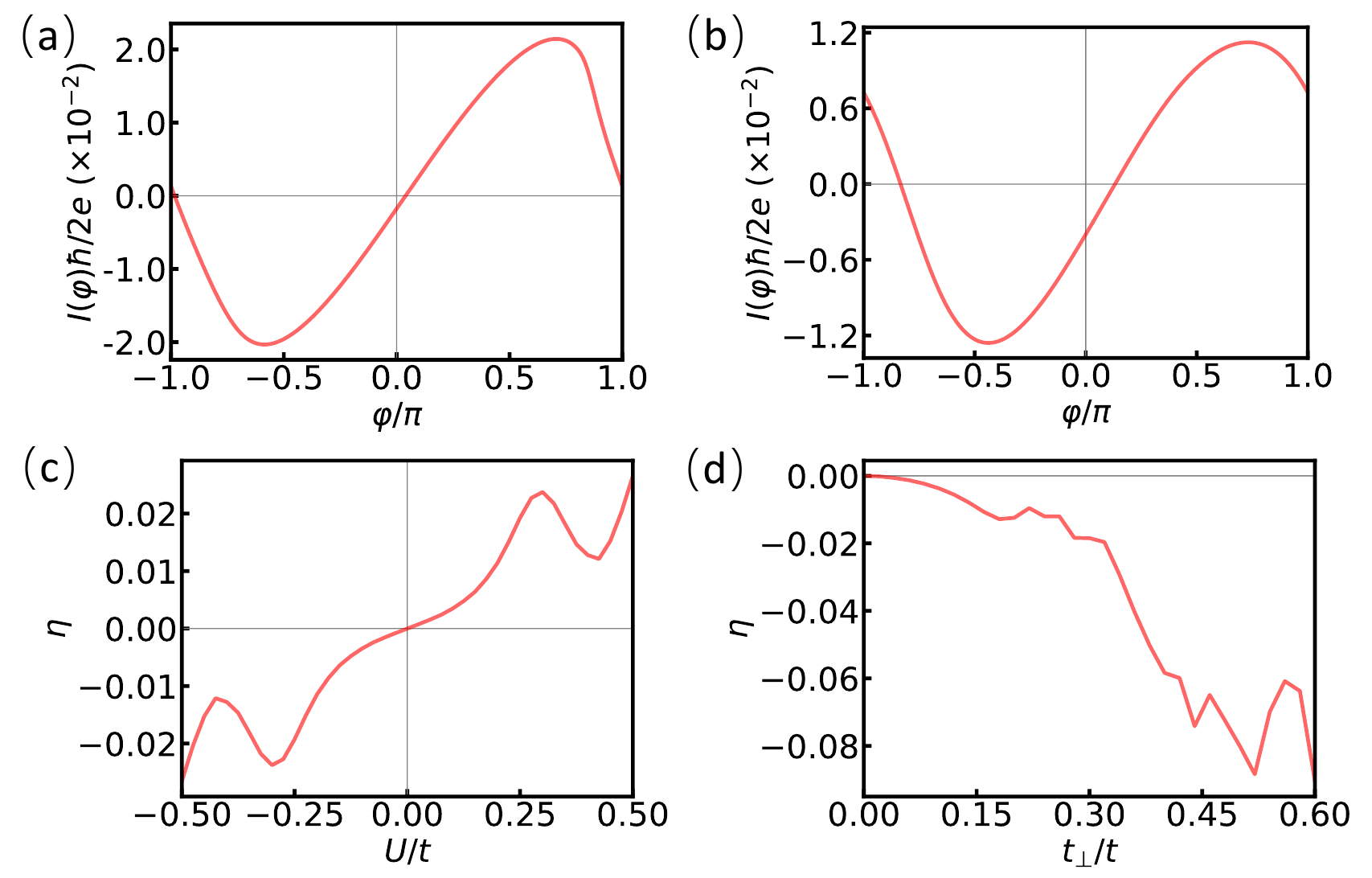}
\caption{The CPR in bilayer zigzag HMs. (a) The JJ is formed by stacking two HMs with the same loop current direction. Inversion symmetry is broken by applying voltage ($U=0.5$) between the two layers. (b) The JJ is formed by stacking two HMs with opposite loop current directions. (c) The diode efficiency $\eta$ as a function of the applied voltage U in bilayer HMs with the same loop current directions. (d) The diode efficiency $\eta$ as a function of the inter-layer hopping $t_\perp$ in bilayer HMs with opposite loop current directions. Parameters are taken as $t=1$, $t_1=0.4t$, $t_\perp=0.4t$, $\phi=\pi/4$, $\Delta=0.2t$, $\mu_0=-0.4t$.}
\label{fig:bilayer_I}
\end{figure}

We next consider the JJs constructed with the bilayer HMs. Based on the analysis of the monolayer case, we can see that breaking the inversion symmetry becomes the key factor in realizing the JDE in the system since the time-reversal symmetry is already broken.
We first consider the case where the orientation of the loop current for the two layers is the same, which still preserves the inversion symmetry and thus prevents the emergence of the JDE. In this system, the inversion symmetry can be broken by applying an external electric field perpendicular to the layers, resulting in a potential difference between the two layers.
In our calculation, this effect is described by a layer-dependent potential term
\begin{equation}
    H_U=
    \frac{U}{2}\sum_{l,i,\sigma} (-1)^l c_{il\sigma}^\dagger c_{il\sigma}
\end{equation}
which is added to the tunneling barrier part of the Hamiltonian in Eq.~\ref{eq:jd_hamilton}.
In this case, the JDE again shows up only for the zigzag JJ.
The CPR for the zigzag JJ with $U=0.5$ is shown in Fig.~\ref{fig:bilayer_I}(a), which clearly shows the JDE induced by the external electric field.
The CPR for the armchair JJ is provided in the Supplemental Material~\cite{sm}.


Besides the case where the inversion symmetry is broken by the external field, we next consider the bilayer magnetoelectric state where the orientations of the loop current of the two layers are opposite to each other so that the inversion symmetry is broken intrinsically.
In this case, the JDE can be realized without the help of an external field as shown in Fig.~\ref{fig:bilayer_I}(b) for the CPR of the zigzag JJ.
The CPR for the armchair JJ again does not show any diode effect~\cite{sm}. 
A thorough symmetry analysis for the absence of the JDE in the armchair JJ is provided in Sec.~\ref{sec:sym}.


\section{Tuning the diode efficiency}
\label{sec:eff}
To investigate the factors influencing the diode effect, we use the diode efficiency factor $\eta$ to quantify the strength of the diode effect. It is defined as:
\begin{equation}
    \eta=\frac{ I_{c}^+ - |I_{c}^-| }{I_{c}^+ + |I_{c}^-|}
    \label{eq:eta}
\end{equation}
where $I_c^+(I_c^-)$ is the critical current along the positive (negative) direction.
For the bilayer HMs, the inversion symmetry is broken by applying the interlayer voltage U for the case with the same loop current direction and the interlayer hopping $t_\perp$ for the case with opposite directions of the loop current breaks the independent $C_2$ rotation symmetry, both of which result in the JDE. Therefore, it is natural to investigate the effects of U and $t_\perp$ on the diode efficiency.

For the zigzag JJ constructed with bilayer HMs with the same loop current direction, we vary the voltage U from -0.5 to 0.5 to investigate its impact on the diode effect. As shown in Fig.\ref{fig:bilayer_I}(c), when the magnitude of U is small, the diode efficiency $\eta$ monotonically increases with U and changes signs as U changes signs.
In this region, the polarity and the strength of the diode efficiency can be well controlled by the external electric field.
This is reasonable since the diode efficiency should increase with the strength of the inversion symmetry breaking as the time-reversal symmetry is fixed by the loop current order and the external voltage is the only source of the inversion symmetry breaking. 
As the magnitude of U further increases, the behavior of the diode efficiency becomes complicated as the band structure of the system is significantly changed by the large value of U.

We next study the effect of the interlayer coupling $t_{\perp}$ on the diode efficiency for the JJ constructed by the bilayer HMs with the opposite loop current directions.   
As shown in Fig.\ref{fig:bilayer_I}(d), the diode effect disappears when $t_{\perp}$ vanishes, which is due to the fact the two layers are decoupled in this limit and each layer has its own $C_2$ rotation symmetry that prevents the emergence of the JDE.
Then, as $t_{\perp}$ increases from 0 to 0.6, the magnitude of $\eta$ monotonically increases with $t_\perp$ when it is weak and starts to exhibit oscillations when it further increases.
This is because $t_{\perp}$ breaks the independent $C_2$ rotation symmetry of the two layers by locking the two layers together keeping the Bernal stacking structure, leading to the emergence of the JDE. Therefore, $t_{\perp}$ plays a similar role here as the voltage U in the case with the same loop current direction for the two layers.

The diode efficiency $\eta$ is sensitive to a range of parameters, including the interlayer voltage $U$, the interlayer coupling $t_{\perp}$, the chemical potential $\mu_0$, the pairing order parameter $\Delta$, the flux phase $\phi$, the size of the system, and so on. Only in the parameter region where the parameters responsible for the symmetry breaking are small, the dependence is smooth and can be understood in the spirit of the perturbation theory.
For the bilayer case, we also explore the effect the chemical potential $\mu_0$ and the pairing amplitude $\Delta$ to $\eta$, in the wide range of $\frac{\Delta}{t}$ and $\frac{\mu_0}{t}$ with the results shown in the Supplemental material~\cite{sm}.
The strong sensitivity of $\eta$ to the parameters requires precise experimental control when implementing the effect in realistic systems.

\section{Symmetry analysis}
\label{sec:sym}
\begin{figure}[h]
\centering
\includegraphics[width=3.5in]{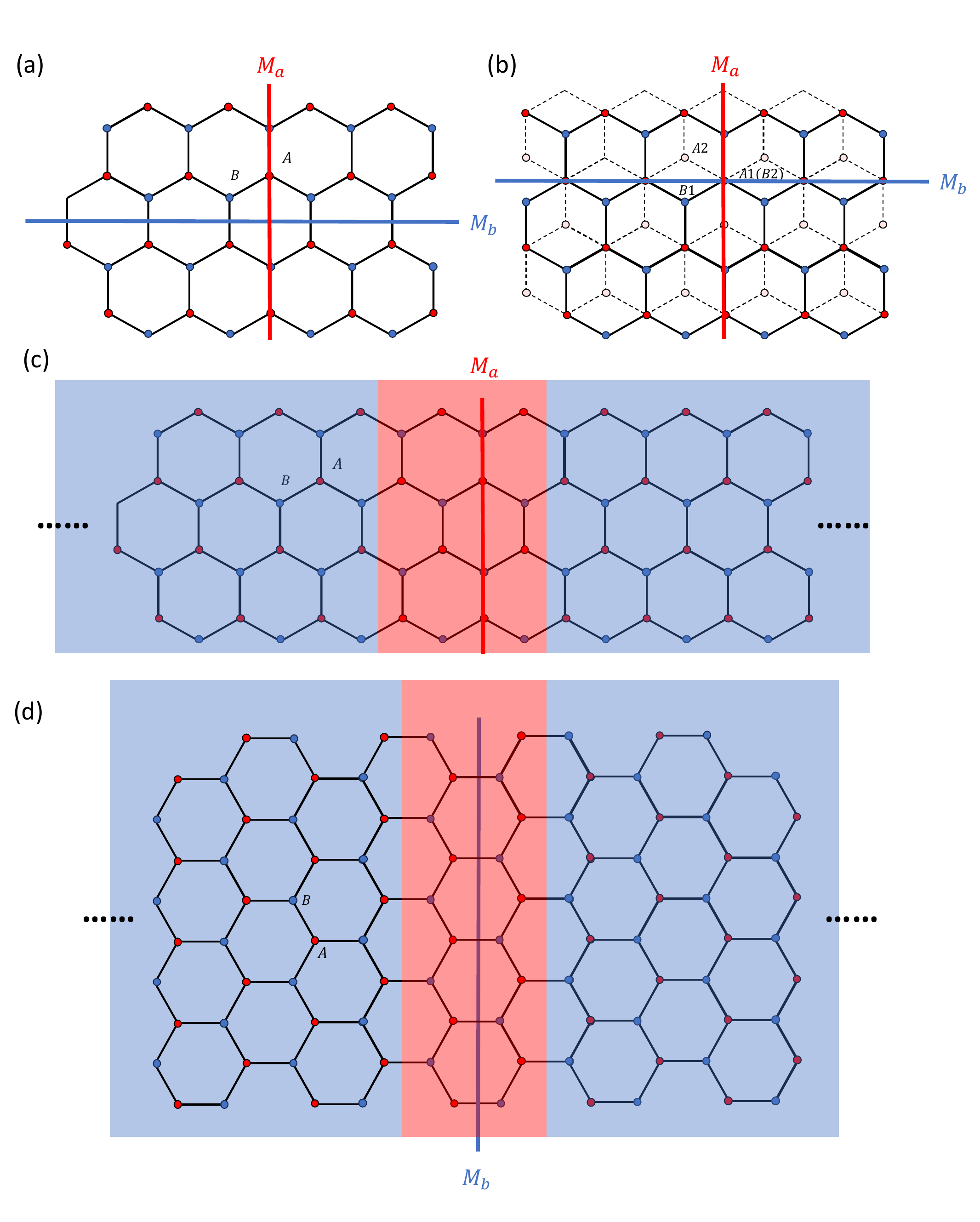}
\caption{Schematics of mirror reflection operations in the monolayer model (a) and the bilayer model (b). For the JJs constructed by the single-layer models, $M_a$ reverses the Josephson current of the zigzag JJs (c), and $M_b$ reverses the Josephson current of the armchair JJs (d). The relations are similar for the JJs constructed by the bilayer model.}
\label{fig:model_symmetry}
\end{figure}

In this section, we demonstrate the reason why the JDE realized in the zigzag JJs disappears in the armchair JJs.
From the symmetry perspective, if the system remains invariant under a certain operation that simultaneously changes the sign of the current operator, it implies that the critical current must have the same magnitude in opposite directions, resulting in the absence of the JDE.
We find that such symmetry operations can be constructed by the following three operations: the mirror reflection ($M_i$), the time-reversal ($\mathcal{T}$), and the $C_2$ rotation about the axis perpendicular to the plane.
Both $C_2$ and $\mathcal{T}$ operations can change the current direction. $M_i$ represents the mirror reflection about the plane perpendicular to the $i$-axis, and the definition of the directions is schematically shown in Fig.~\ref{fig:model_symmetry}, where $i=a$ corresponds to the zigzag direction and $i=b$ represents the armchair direction. For the zigzag JJ, $M_a$ changes the current direction while $M_b$ does not. Conversely, for the armchair JJ, $M_b$ changes the current direction, but $M_a$ does not. 

For JJs constructed by different models with different junction directions, the possible symmetry operations that prevent the emergence of the JDE are summarized in Table.~\ref{table:Symmetry analysis}. For example, for the armchair JJ constructed by the monolayer HM model with staggered lattice potential, operations $M_b$, $C_2$ and $\mathcal{T}$ all change the direction of the current. Therefore, the combined operation $M_b C_2 \mathcal{T}$ changes the direction of the current while keeping the system invariant. Hence, the JDE is absent in the armchair JJ constructed by the HM model with staggered lattice potential. For zigzag JJ constructed by the same model, we can not find any symmetry operations that also change the direction of the current which is represented by the `$\times$' sign in the table.

\begin{table}[h!]
\centering
\begin{tabular}{|c|c|c|c|}
\hline
\textcolor{blue}{monolayer I} & \textcolor{blue}{operations} & \textcolor{blue}{JDE} & \textcolor{blue}{control strategy}\\
\hline
armchair HM & $C_2$ & No & \multirow{4}{*}{/}\\
armchair MHM  & $M_b$ & No &\\
zigzag HM & $C_2$ & No & \\
zigzag MHM  & $\times$ & Yes & \\
\hline
\textcolor{blue}{monolayer II} & \textcolor{blue}{operations} & \textcolor{blue}{JDE} & \textcolor{blue}{control strategy}\\
\hline
armchair HM & $M_b C_2 \mathcal{T}$ & No & \multirow{4}{*}{\makecell{place the system \\on a substrate of h-BN}}\\
armchair MHM  & $M_b C_2 \mathcal{T}$ & No &\\
zigzag HM & $\times$ & Yes & \\
zigzag MHM  & $\times$ & Yes & \\
\hline
\textcolor{blue}{bilayer I} & \textcolor{blue}{operations} & \textcolor{blue}{JDE} & \textcolor{blue}{control strategy} \\
\hline
armchair HMs   & $M_b C_2 \mathcal{T}$ & No &  \multirow{2}{*}{\makecell{apply an external electric field \\ perpendicular to the barrier}}\\
zigzag HMs  & $\times$ & Yes & \\
\hline
\textcolor{blue}{bilayer II} & \textcolor{blue}{operations} & \textcolor{blue}{JDE} & \textcolor{blue}{control strategy}\\
\hline
armchair HMs & $M_b C_2 \mathcal{T}$ & No & \multirow{2}{*}{/}\\
zigzag HMs & $\times$ & Yes & \\
\hline
\end{tabular}
\caption{Symmetry analysis for the JDE. The first column lists all the models used to construct the Josephson junction, where monolayer I(II) corresponds to models without(with) staggered potential, bilayer I corresponds to bilayer models with the same current direction and interlayer potential, and bilayer II corresponds to bilayer models with opposite current direction. The second column lists the symmetry operations that prevent the emergence of the JDE where the `$\times$' sign means no such symmetry operation. The third column shows whether certain Josephson junctions can show the JDE or not. The last column describes the possible experimental techniques for tuning the model parameters.}
\label{table:Symmetry analysis}
\end{table}

\section{The impact of disorder}
\label{sec:dis}
\begin{figure}
\centering
\includegraphics[width=3.5in]{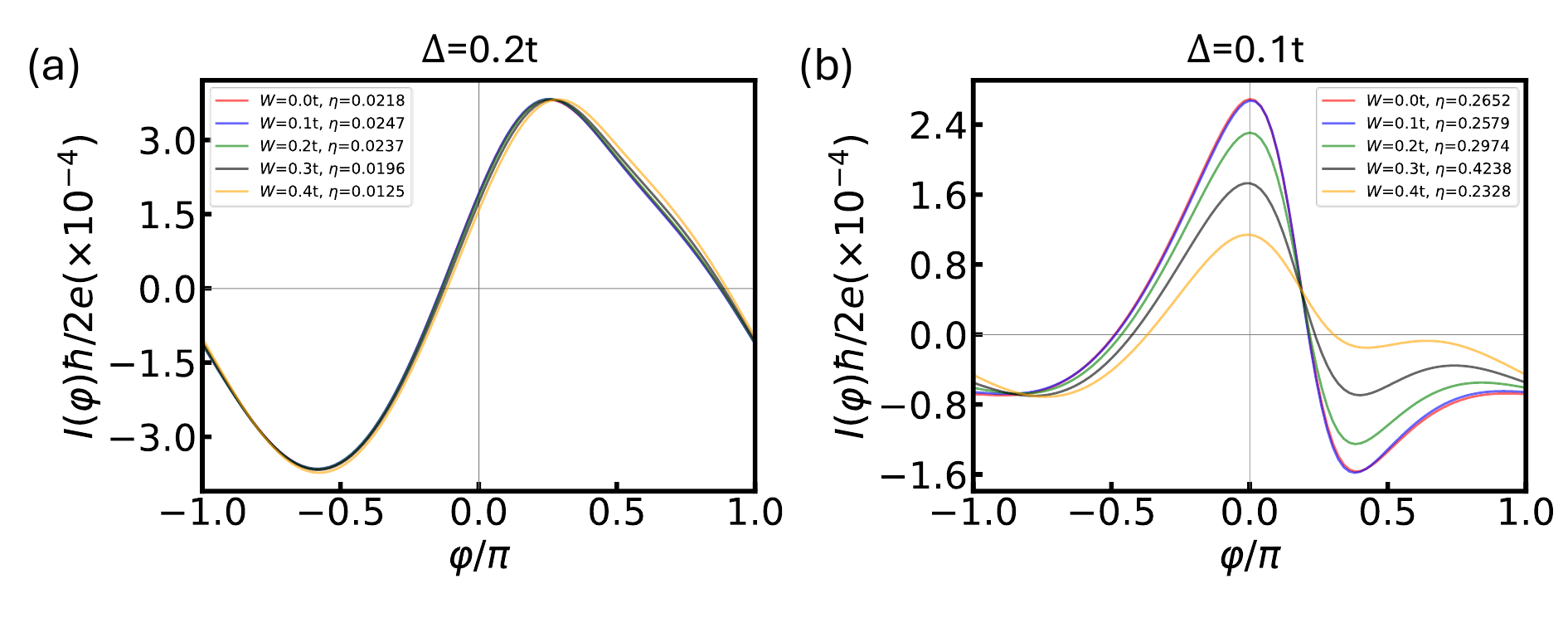}
\caption{The CPR and the diode efficiency $\eta$ of the zigzag JJs constructed by the single-layer MHM for various disorder strength W. (a) When $\Delta=0.2t$, the three CPR curves for  $W=0t, 0.1t, 0.2t$ overlap due to the minimal offsets. (b) When $\Delta=0.1t$, the current decreases with increasing W, while the diode effect remains stable. Other parameters are taken as $t=1$, $t_1=0.4t$, $\phi=\pi/4$, $\mu_0=0$.}
\label{fig:disorder}
\end{figure}

To investigate the impact of impurities on the diode effect, we study the JDE with a random on-site potential describing the disorder effect introduced by the impurities.
We use the random potentials obeying the uniform distribution ranging from $-\frac{W}{2}$ to $\frac{W}{2}$ and the results are averaged over 100 disorder configurations. 
Since the translation invariance along the $y$-direction is broken by the random impurity potential, we perform the calculation in real space, which significantly increases the computational complexity.
We therefore reduce the model size. The size in the $x$-direction remains unchanged, while in the $y$-direction, it is reduced from 100 unit cells (4$\times$100 atoms) to 10 unit cells (4$\times$10 atoms).  
The results for the zigzag JJ constructed by the single-layer MHM are shown in Fig.~\ref{fig:disorder}, where the averaged CPR and diode efficiency $\eta$ obtained from the disordered model with various disorder strengths $W$ are compared to those of the impurity-free model. 

We perform the calculation for two values of the superconducting order parameter with $\Delta$=$0.2t$ and $0.1t$. The results clearly show that the CPRs for the disordered JJs still show the diode effect and the diode efficiency $\eta$ persists with the increase of the disorder strength. 
Moreover, for the relatively strong pairing strength $\Delta=0.2t$, the critical currents are less affected by the disorder strength $W$ as evidenced by the close resemblance of the CPRs across different disorder strengths, leading to a stable diode efficiency $\eta$ for $W\leq 0.2t$. It starts to decrease slightly for $W$ above $0.1t$. On the other hand, for the relatively weak pairing strength $\Delta=0.1t$, the supercurrents become weaker and more susceptible to disorder, which manifests as a rapid decay of the critical supercurrents in both directions as shown in Fig.~\ref{fig:disorder}(b). However, since the breaking of time-reversal and inversion symmetries is governed by the loop currents rather than $\Delta$, the diode effect become more pronounced in this case and $\eta$ starts to decrease for $W$ above $0.3t$. This rules out the possibility that the “critical” disorder strength where $\eta$ starts to decrease is restricted by $\Delta$, thereby indicating that the diode effect is robust against the disorder.

The robustness of the JDE against disorder can also be understood as follows. The diode efficiency $\eta$ defined in Eq.~\ref{eq:eta} quantifies the relative difference in critical supercurrents flowing in opposite directions. Its physical origin lies in the breaking of certain symmetries, such as time-reversal and inversion symmetries, that otherwise relate the current in opposite directions. In this sense, $\eta$ serves as a measure of symmetry breaking. Consequently, it is reasonable to expect that $\eta$ remains robust when the disorder is not too strong, as it cannot restore the broken symmetries responsible for the diode effect.


\section{Discussion}
\label{sec:con}
In summary, we have demonstrated that the JDE can be realized in the JJs constructed by the monolayer and bilayer honeycomb systems with loop current order, which directly breaks the time-reversal symmetry.
For the monolayer system, the inversion symmetry can be broken by either applying a staggered sublattice potential for the HM or by reverting parts of the loop current directions to form the MHM, leading to the JDE. 
For the bilayer system, the JDE can be realized by applying an external electric field perpendicular to the system resulting in an electrically controllable Josephson diode, or by stacking the two layers of HMs with opposite loop current directions, which leads to an intrinsic field-free Josephson diode.
Moreover, we also find that the JDE disappears when the JJ is constructed along the armchair edge which has an extra symmetry to prevent the emergence of the JDE.
Finally, we demonstrate that the JDE is robust against disorder induced by impurity by calculating the average CPR for the systems with random onsite potential.
The JDE persists in the average CPR over 100 random configurations. While this work focuses on the short-junction limit, we expect the JDE to persist in longer junctions, since the symmetry-breaking pattern remains unchanged as the junction size increases.

We expect this mechanism to realize the JDE can be easily applied to other hexagonal systems such as the Kagome systems. 
One of the possible platforms to realize this field-free JDE induced by the loop current is the Kagome metal system, which hosts a special charge density wave state as a promising candidate for the loop current order~\cite{cdw5,cdw6,vidya_2024}.
We hope it stimulates the exploration of the field-free JDE and expands the scope of the possible material systems to realize this effect.



\begin{acknowledgments}
This work is supported by the National Natural Science Foundation of China (NSFC) Grants No. 12274279. We gratefully acknowledge HZWTECH for providing computation facilities.
\end{acknowledgments}

\bibliography{jose1}

\clearpage
\onecolumngrid
\begin{center}
\textbf{\large Supplemental Material: Josephson diodes induced by the loop current states}
\end{center}

\setcounter{equation}{0}
\setcounter{figure}{0}
\setcounter{table}{0}
\setcounter{page}{1}
\makeatletter
\renewcommand{\theequation}{S\arabic{equation}}
\renewcommand{\thefigure}{S\arabic{figure}}
\renewcommand{\bibnumfmt}[1]{[S#1]}
\renewcommand{\citenumfont}[1]{S#1}

\onecolumngrid
\section{Construction of the Hamiltonian of a Josephson junction}
\begin{figure}[h]
\centering
\includegraphics[width=4in]{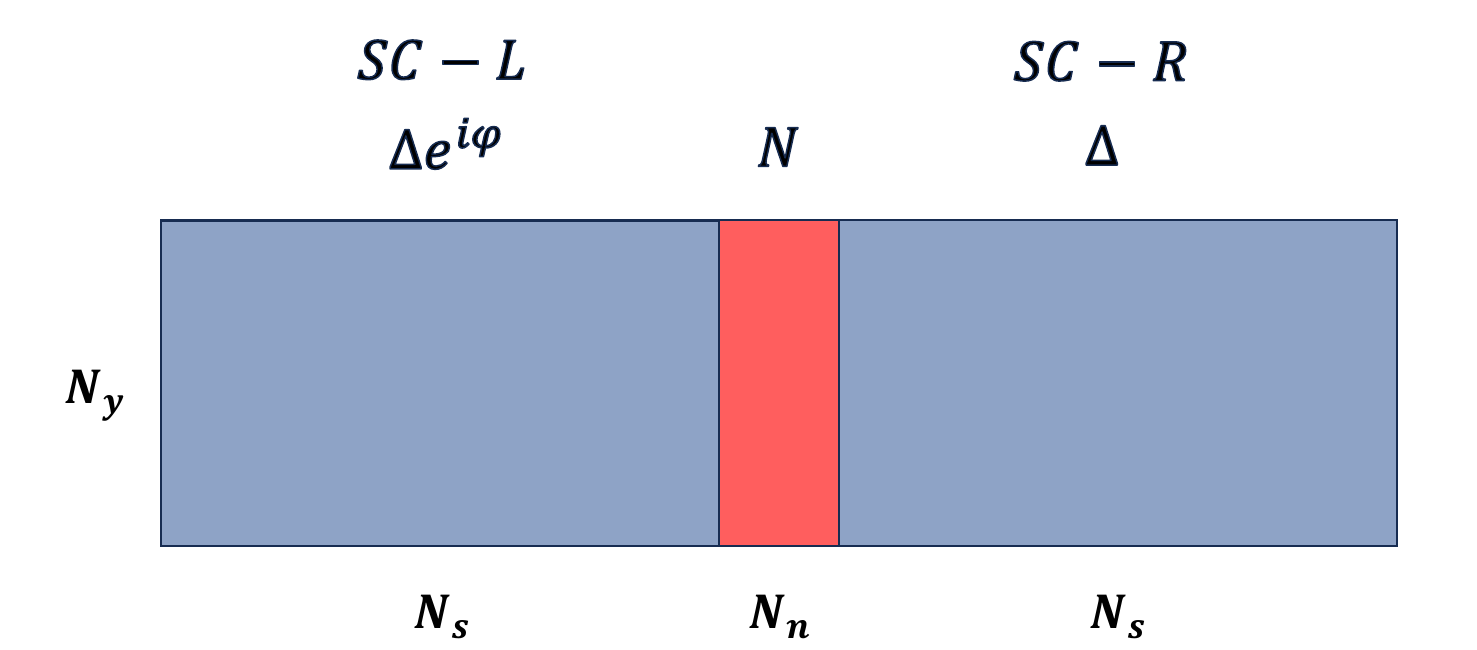}
\caption{Schematic of a JJ. The lengths of the left and the right superconductors are both $N_s$ and the middle non-superconducting region is $N_n$. The width of the junction is $N_y$ unit cells. The superconducting phase bias $\varphi$ for the JJ is assigned to the left superconducting region, with the phase of the right superconducting region set to 0.}
\label{fig:junction_2D}
\end{figure}

We study a planar Josephson junction (JJ) in which the width of the junction spans $N_y$ unit cells. The two superconducting regions each have a length of $N_s$ unit cells, while the central non-superconducting region extends over $N_n$ unit cells as shown in Fig.~\ref{fig:junction_2D}.
Thus the Hamiltonian of the JJ can be expressed as 
\begin{equation}
    \begin{split}
        H&=H_{SL}+H_{SN}+H_{N}+H_{NS}+H_{SR}\\
        H_{SL}&=H_{0}+\Delta e^{i\varphi} \sum_i c_{i\uparrow}^\dagger c_{i\downarrow}^\dagger+h.c.-\mu_0 \sum_{i,\sigma} c_{i \sigma}^\dagger c_{i \sigma}\\
        H_N&=H_{0}-\mu_0 \sum_{i,\sigma} c_{i \sigma}^\dagger c_{i \sigma}\\
        H_{SR}&=H_{0}+\Delta \sum_i c_{i\uparrow}^\dagger c_{i\downarrow}^\dagger+h.c.-\mu_0 \sum_{i,\sigma} c_{i \sigma}^\dagger c_{i \sigma}\\
        H_{SN}&=-t \sum_{\left \langle i,j \right\rangle\sigma , i\in SL, j\in N} c_{i \sigma}^\dagger c_{j \sigma} + t_1 \sum_{\left \langle\left \langle i,j \right\rangle\right \rangle\sigma , i\in SL, j\in N} e^{-i\nu_{ij}\phi} c_{i\sigma}^\dagger c_{j\sigma} +h.c. \\
        H_{NS}&=-t \sum_{\left \langle i,j \right\rangle\sigma , i\in N, j\in SR} c_{i \sigma}^\dagger c_{j \sigma} + t_1 \sum_{\left \langle\left \langle i,j \right\rangle\right \rangle\sigma , i\in N, j\in SR} e^{-i\nu_{ij}\phi} c_{i\sigma}^\dagger c_{j\sigma} +h.c.
    \end{split}
    \label{eq:jd_hamilton}
\end{equation}
where $H_{0}$ represents the Hamiltonian of monolayer ($H_m$) or bilayer ($H_b$) system.
For the monolayer system, it can be written as
\begin{equation}
    \begin{split}
        H_m=-t\sum_{\left \langle i,j \right\rangle\sigma}c_{i\sigma}^\dagger c_{j\sigma}-
        t_1 \sum_{\left \langle\left \langle i,j \right\rangle\right \rangle\sigma} e^{-i\nu_{ij}\phi} c_{i\sigma}^\dagger c_{j\sigma}
    \end{split}
    \label{con:monolayer_H}
\end{equation}
describing the monolayer Haldane or modified Haldane models.
For the bilayer system, it can be expressed as
\begin{equation}
    \begin{split}
        H_b&=H_1+H_2+H_{12}\\
        H_l&=-t\sum_{\left \langle i,j \right\rangle\sigma}c_{il\sigma}^\dagger c_{jl\sigma}-t_1 \sum_{\left \langle\left \langle i,j\right\rangle\right\rangle\sigma} e^{-i\nu_{lij}\phi} c_{il\sigma}^\dagger c_{jl\sigma}\\
        H_{12}&=-t_\perp \sum_{i,\sigma} c_{i,A1 \sigma}^\dagger c_{i,B2 \sigma}+h.c.
    \end{split}
    \label{con:bilayer_H}
\end{equation}
describing the bilayer Haldane or modified Haldane model coupled through the interlayer coupling $t_{\perp}$.
Thus, $H_{SL}$ and $H_{SR}$ correspond to the Hamiltonian of the superconducting region on the left and right sides. $H_{N}$ describes the normal region in the middle with the superconducting pairing term set to 0. 
$H_{SN}$ and $H_{NS}$ correspond to the spin-conserving hoppings between the superconducting and normal sections which include both the nearest-neighbor and next-nearest-neighbor hoppings and take the same values and patterns as those in the normal region. The superconducting phase bias for the JJ is assigned to the left superconducting region as $\varphi$ so that the phase in the right superconducting region is set to $0$. The amplitude of the order parameter is taken as $\Delta=0.2t$ for all calculations in this paper.   $\mu_0$ represents the chemical potential of the entire material, which is set to -0.4t in the bilayer model, and set to 0 in the monolayer system.
Moreover, a staggered lattice potential can be described by an extra term $H_s=V_s\sum_{i,\sigma} (c_{i,A \sigma}^\dagger c_{i,A \sigma}-c_{i,B \sigma}^\dagger c_{i,B \sigma})$ added into the monolayer Hamiltonian and an out-of-plane electric field for the bilayer system can be described by a term $H_U=U/2\sum_{l,i,\sigma} (-1)^l c_{il\sigma}^\dagger c_{il\sigma}$.
The periodic boundary condition is adopted along the width of the junction and the total Hamiltonian is diagonalized numerically. 
Then the Josephson currents can be calculated as
\begin{equation}
    \begin{split}
        I(\varphi)=\frac{2e}{\hbar}\partial_\varphi \sum_n f(\epsilon_n(\varphi))\epsilon_n(\varphi)
    \end{split}
\end{equation}
with $\epsilon_n(\varphi)$ the nth eigenvalue for $H(\varphi)$ and $f(\epsilon)$ the Fermi distribution function.

\section{CPR for the JJ constructed by the monolayer HM model}
\begin{figure}[h]
\centering
\includegraphics[width=6.5in]{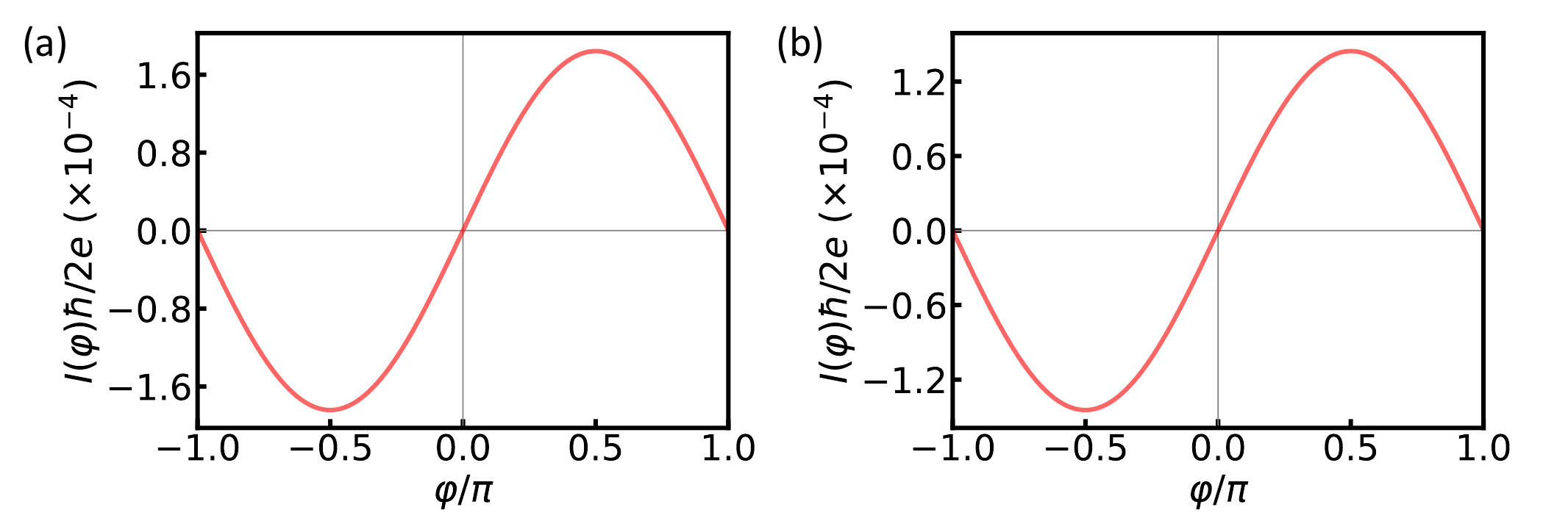}
\caption{The CPR of the JJ constructed by the monolayer HM. (a) The JJ is along the zigzag edge direction and (b) the JJ is along the armchair edge direction. Parameters are taken as $t=1$, $t_1=0.4t$, $\phi=\pi/4$, $\Delta=0.2t$, $\mu_0=0$.}
\label{fig:monolayer_hm}
\end{figure}

For the JJ constructed by the monolayer HM model, the CPR shows no diode effect since the HM model has the inversion symmetry ($C_2$ for the monolayer system) as shown in Fig.~\ref{fig:monolayer_hm}.

\section{CPR for the armchair JJ constructed by the bilayer models}
\begin{figure}[h]
\centering
\includegraphics[width=6.5in]{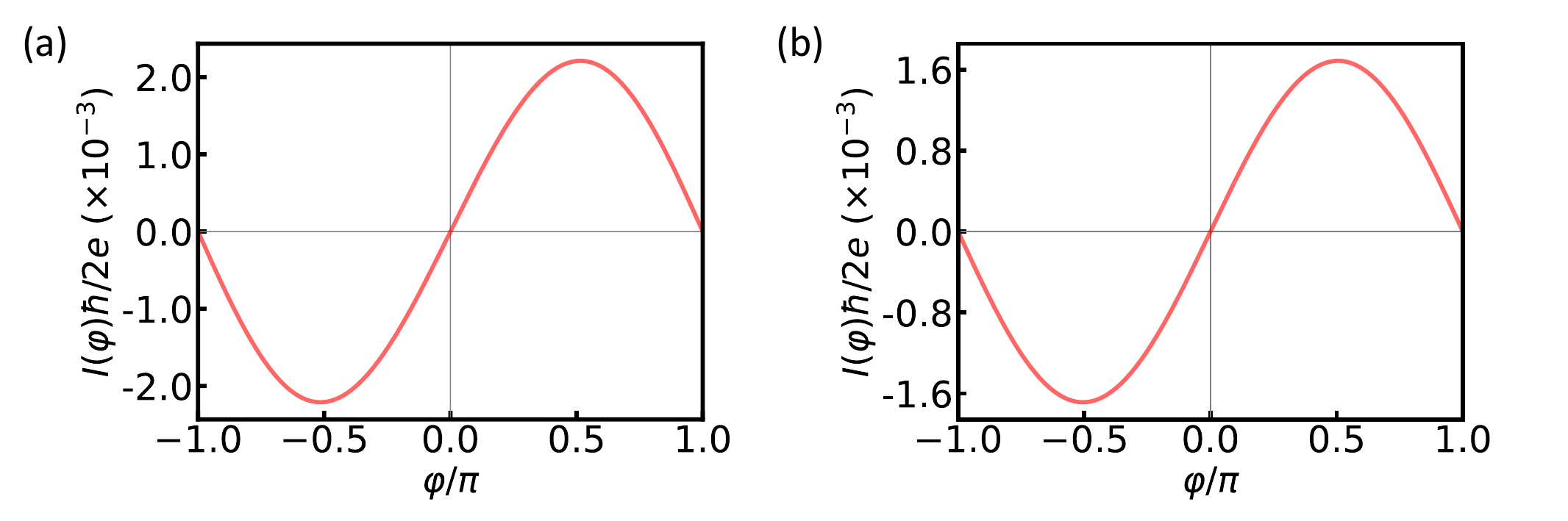}
\caption{The CPR in bilayer armchair HMs. (a) The JJ is formed by stacking two HMs with the same loop current direction and an external electric field perpendicular to the system is applied. (b) The JJ is formed by stacking two HMs with opposite loop current directions. Parameters are taken as $t=1$, $t_1=0.4t$, $t_\perp=0.4t$, $\phi=\pi/4$, $\Delta=0.2t$, $\mu_0=-0.2t$ and $U=0.5t$ (for (a) only).}
\label{fig:bilayer_armchair}
\end{figure}

As mentioned in the main text, the JDE can be achieved by applying an out-of-plane electric field to the bilayer HMs with the same loop current direction or stacking the two layers of the HMs with opposite loop current directions in the zigzag JJs.
However, if the JJs are constructed along the armchair edge, the JDE for both cases disappears, which is related to the mirror symmetry of the armchair JJs. Here, we provide the numerical evidence to verify these results in Fig.~\ref{fig:bilayer_armchair}, where the Josephson current of the armchair JJ $I(\varphi)$ is an odd function of the phase bias $\varphi$ and thus does not show any diode effect.

\section{The effect of the chemical potential $\mu_0$ and pairing strength $\Delta$}
\begin{figure}[h]
\centering
\includegraphics[width=6.5in]{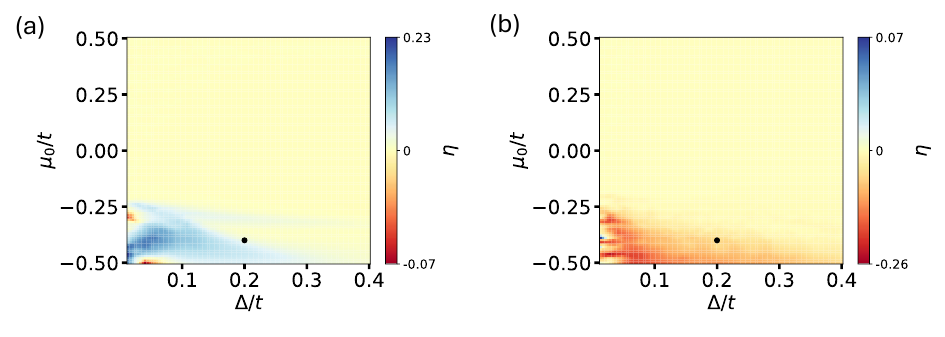}
\caption{The calculated diode efficiency $\eta$ for different values of $\mu_0$ and $\Delta$ for JJs constructed by the bilayer HM. The red region indicates $\eta$ is negative, the blue region indicates $\eta$ is positive, and the yellow region represents regions where the JDE is not significant but still present. The black dots mark the parameters selected in the main text, as in Fig.4(a)(b). (a) The JJ is formed by stacking two HMs with the same loop current direction. Inversion symmetry is broken by applying voltage ($U=0.5t$) between the two layers. (b) The JJ is formed by stacking two HMs with opposite loop current directions. Parameters are taken as $t=1$, $t_1=0.4t$, $t_\perp=0.4t$, $\phi=\pi/4$.}
\label{fig:mu_delta}
\end{figure}

To explore the effect of the chemical potential $\mu_0$ and the pairing amplitude $\Delta$ to the diode efficiency $\eta$ for the JJs constructed by the bilayer HM, we calculate $\eta$ for various values of $\frac{\Delta}{t}$ and $\frac{\mu_0}{t}$. The pairing amplitude $\Delta$ is varied from $0$ to $0.4t$ and the chemical potential $\mu_0$ is varied from $-0.5t$ to $0.5t$ in the bilayer model with other parameters unchanged. The results are shown in Fig.~\ref{fig:mu_delta} where
the parameters used in the paper are marked with black dots. We can see that the diode efficiency $\eta$ increases significantly when $\mu_0<-0.3t$ and weak $\Delta$, while it remains relatively low in other parameter regions. 

\end{document}